\documentclass[twocolumn]{aastex631}
\submitjournal{ApJ Letters}

\usepackage{textcomp}
\usepackage{graphicx}

\UseRawInputEncoding

\begin{document}

\title{GOALS-JWST: Pulling Back the Curtain on the AGN and Star Formation in VV 114}

\author[0000-0002-5807-5078]{J. Rich}
\affiliation{The Observatories of the Carnegie Institution for Science, 813 Santa Barbara Street, Pasadena, CA 91101}

\author[0000-0002-5828-7660]{S. Aalto}
\affiliation{Department of Space, Earth and Environment, Chalmers University of Technology, 412 96 Gothenburg, Sweden}

\author[0000-0003-2638-1334]{A.S. Evans}
\affiliation{National Radio Astronomy Observatory, 520 Edgemont Rd, Charlottesville, VA, 22903, USA}
\affiliation{Department of Astronomy, University of Virginia, 530 McCormick Road, Charlottesville, VA 22903, USA}

\author[0000-0002-2688-1956]{V. Charmandaris}
\affiliation{Department of Physics, University of Crete, Heraklion, 71003, Greece}
\affiliation{Institute of Astrophysics, Foundation for Research and Technology-Hellas (FORTH), Heraklion, 70013, Greece}
\affiliation{School of Sciences, European University Cyprus, Diogenes street, Engomi, 1516 Nicosia, Cyprus}

\author[0000-0003-3474-1125]{G. C. Privon}
\affiliation{National Radio Astronomy Observatory, 520 Edgemont Rd, Charlottesville, VA, 22903, USA}
\affiliation{Department of Astronomy, University of Florida, P.O. Box 112055, Gainesville, FL 32611, USA}

\author[0000-0001-8490-6632]{T. Lai}
\affiliation{IPAC, California Institute of Technology, 1200 E. California Blvd., Pasadena, CA 91125, USA}

\author[0000-0003-4268-0393]{H. Inami}
\affiliation{Hiroshima Astrophysical Science Center, Hiroshima University, 1-3-1 Kagamiyama, Higashi-Hiroshima, Hiroshima 739-8526, Japan}

\author[0000-0002-1000-6081]{S. Linden}
\affiliation{Department of Astronomy, University of Massachusetts at Amherst, Amherst, MA 01003, USA}

\author[0000-0003-3498-2973]{L. Armus}
\affiliation{IPAC, California Institute of Technology, 1200 E. California Blvd., Pasadena, CA 91125, USA}

\author[0000-0003-0699-6083]{T. Diaz-Santos}
\affiliation{Institute of Astrophysics, Foundation for Research and Technology-Hellas (FORTH), Heraklion, 70013, Greece}
\affiliation{School of Sciences, European University Cyprus, Diogenes street, Engomi, 1516 Nicosia, Cyprus}

\author[0000-0002-7607-8766]{P. Appleton}
\affiliation{IPAC, California Institute of Technology, 1200 E. California Blvd., Pasadena, CA 91125, USA}

\author[0000-0003-0057-8892]{L. Barcos-Mu\~noz}
\affiliation{Department of Astronomy, University of Virginia, 530 McCormick Road, Charlottesville, VA 22903, USA}

\author[0000-0002-5666-7782]{T. B\"oker}
\affiliation{European Space Agency, Space Telescope Science Institute, Baltimore, MD 21218, USA}

\author[0000-0003-3917-6460]{K. L. Larson}
\affiliation{AURA for the European Space Agency (ESA), Space Telescope Science Institute, 3700 San Martin Drive, Baltimore, MD 21218, USA}

\author[0000-0002-9402-186X]{D. R.~Law}
\affiliation{Space Telescope Science Institute, 3700 San Martin Drive, Baltimore, MD 21218, USA}

\author[0000-0001-6919-1237]{M. A. Malkan}
\affiliation{Department of Physics \& Astronomy, 430 Portola Plaza, University of California, Los Angeles, CA 90095, USA}

\author[0000-0001-7421-2944]{A. M. Medling}
\affiliation{Department of Physics \& Astronomy and Ritter Astrophysical Research Center, University of Toledo, Toledo, OH 43606,USA}
\affiliation{ARC Centre of Excellence for All Sky Astrophysics in 3 Dimensions (ASTRO 3D); Australia}

\author[0000-0002-3139-3041]{Y. Song}
\affiliation{Department of Astronomy, University of Virginia, 530 McCormick Road, Charlottesville, VA 22903, USA}
\affiliation{National Radio Astronomy Observatory, 520 Edgemont Rd, Charlottesville, VA, 22903, USA}

\author[0000-0002-1912-0024]{V. U}
\affiliation{Department of Physics and Astronomy, 4129 Frederick Reines Hall, University of California, Irvine, CA 92697, USA}

\author[0000-0001-5434-5942]{P. van der Werf}
\affiliation{Leiden Observatory, Leiden University, PO Box 9513, 2300 RA Leiden, The Netherlands}

\author{T. Bohn}
\affiliation{Hiroshima Astrophysical Science Center, Hiroshima University, 1-3-1 Kagamiyama, Higashi-Hiroshima, Hiroshima 739-8526, Japan}

\author[0000-0002-1207-9137]{M. J. I. Brown}
\affiliation{School of Physics and Astronomy, Monash University, Clayton, VIC 3800, Australia}

\author[0000-0002-1392-0768]{L. Finnerty}
\affiliation{Department of Physics \& Astronomy, 430 Portola Plaza, University of California, Los Angeles, CA 90095, USA}

\author[0000-0003-4073-3236]{C. Hayward}
\affiliation{Center for Computational Astrophysics, Flatiron Institute, 162 Fifth Avenue, New York, NY 10010, USA}

\author[0000-0001-6028-8059]{J. Howell}
\affiliation{IPAC, California Institute of Technology, 1200 E. California Blvd., Pasadena, CA 91125}

\author[0000-0002-4923-3281]{K. Iwasawa}
\affiliation{Institut de Ci\`encies del Cosmos (ICCUB), Universitat de Barcelona (IEEC-UB), Mart\'i i Franqu\`es, 1, 08028 Barcelona, Spain}
\affiliation{ICREA, Pg. Llu\'is Companys 23, 08010 Barcelona, Spain}

\author[0000-0003-2743-8240]{F. Kemper}
\affiliation{Institut de Ciencies de l'Espai (ICE, CSIC), Can Magrans, s/n, 08193 Bellaterra, Barcelona, Spain}
\affiliation{ICREA, Pg. Llu\'is Companys 23, 08010 Barcelona, Spain}
\affiliation{Institut d'Estudis Espacials de Catalunya (IEEC), E-08034 Barcelona, Spain}

\author[0000-0001-7712-8465]{J. Marshall}
\affiliation{Glendale Community College, 1500 N. Verdugo Rd., Glendale, CA 91208}

\author[0000-0002-8204-8619]{J. M. Mazzarella}
\affiliation{IPAC, California Institute of Technology, 1200 E. California Blvd., Pasadena, CA 91125}

\author[0000-0002-6149-8178]{J. McKinney} 
\affiliation{Department of Astronomy, University of Massachusetts, Amherst, MA 01003, USA.}

\author[0000-0002-2713-0628]{F. Muller-Sanchez}
\affiliation{Department of Physics and Materials Science, The University of Memphis, 3720 Alumni Avenue, Memphis, TN 38152, USA}

\author[0000-0001-7089-7325]{E.J.\,Murphy}
\affiliation{National Radio Astronomy Observatory, 520 Edgemont Road, Charlottesville, VA 22903, USA}

\author[0000-0002-1233-9998]{D. Sanders}
\affiliation{Institute for Astronomy, University of Hawaii, 2680 Woodlawn Drive, Honolulu, HI 96822}

\author{B. T. Soifer}
\affiliation{Division of Physics, Mathematics and Astronomy, California Institute of Technology, 1200 E. California Blvd., Pasadena, CA 91125}

\author[0000-0002-2596-8531]{S. Stierwalt}
\affiliation{Physics Department, 1600 Campus Road, Occidental College, Los Angeles, CA 90041, USA}

\author[0000-0001-7291-0087]{J. Surace}
\affiliation{IPAC, California Institute of Technology, 1200 E. California Blvd., Pasadena, CA 91125}

\begin{abstract}
We present results from the James Webb Space Telescope (\emph{JWST}) Director's Discretionary Time Early Release Science (ERS) program 1328 targeting the nearby, Luminous Infrared Galaxy (LIRG), VV 114. We use the MIRI and NIRSpec instruments to obtain integral-field spectroscopy of the heavily obscured Eastern nucleus (V114E) and surrounding regions. The spatially resolved, high-resolution, spectra reveal the physical conditions in the gas and dust over a projected area of 2-3 kpc that includes the two brightest IR sources, the NE and SW cores. Our observations show for the first time spectroscopic evidence that the SW core hosts an AGN as evidenced by its very low 6.2$\mu$m and 3.3$\mu$m PAH equivalent widths (0.12 and 0.017 $\mu$m respectively) and mid and near-IR colors. Our observations of the NE core show signs of deeply embedded star formation including absorption features due to aliphatic hydrocarbons, large quantities of amorphous silicates, as well as HCN due to cool gas along the line of sight. We detect elevated [Fe II]/Pf$\alpha$ consistent with extended shocks coincident with enhanced emission from warm H$_{2}$, far from the IR-bright cores and clumps. We also identify broadening and multiple kinematic components in both H$_{2}$ and fine structure lines caused by outflows and previously identified tidal features.
\end{abstract}

\section{Introduction}
VV 114 (Arp236, IC1623) is an interacting system undergoing vigorous starburst activity. With an infrared luminosity of L$_{IR} \sim 4.5 \times 10^{11}$ L$_{\odot}$, and a distance of 80 Mpc, it is one of the brightest objects in the IRAS Bright Galaxy Sample \citep{Soifer87}. It appears to be an early-stage merger of two galaxies that are aligned east-west with a projected nuclear separation of $\sim$8 kpc, designated in the literature as VV 114E and VV 114W.  At optical wavelengths, VV 114 shows a highly disturbed morphology with very faint tidal tails extending over 25 kpc from the center \citep{Arp66}. The western component, VV 114W, is more extended than the eastern one, and dominates the emission at short wavelengths. Much of the mid-infrared emission is diffuse and extended over several kpc with some indication of an AGN based on the mid-infrared color of the more compact nuclear region in VV 114E \citep{Lefloch02}. ALMA observations show abundant cold, dense gas (traced by e.g. CO, HCO, HCN), evidence for shocked gas in the overlap region between the two galaxies (traced by Methanol), a molecular outflow, and a possible buried AGN in VV 114E \citep{Iono13,Saito15,Saito17,Saito18}. The majority of the IR emission and by extension total energy output of the system is dominated by VV 114E and the extreme UV/optical to IR ratios of VV 114 make it a more plausible analog to high-z IR luminous mergers \citep{Charmandaris04,Howell10}.

Mid-infrared spectra centered on VV 114E taken with \emph{Spitzer} IRS ($\sim10\arcsec \times 36\arcsec$) show a moderately strong 9.7$\mu$m Silicate absorption (s$_{9.7 \mu m} = -0.98$), a 6.2$\mu$m PAH equivalent width intermediate to LIRGs dominated by star formation or AGN (EQW$_{6.2 \mu m} = 0.30 \mu m$), and a ratio of H$_{2}$ to PAH luminosity slightly above values associated with photodissociation region emission \citep{Guillard12a,Stierwalt13,Stierwalt14}. No coronal lines (e.g. [Ne V]) were detected in the Spitzer spectra, and fine structure line flux ratios were consistent with LIRGs primarily dominated by star formation with some composite AGN/starburst activity \citep{Inami13}. Observations with Chandra, XMM, and NuSTAR of VV 114E indicate that the X-ray emission appears to be generated primarily through star fomation, with little to no X-ray emission coming from an AGN \citep{Grimes06,Garofali20,Ricci21}. Visible wavelength integral field spectroscopy indicates a mix of star formation and shock emission, the latter indicated by elevated emission line ratios and line widths across both galaxies \citep{Rich11,Rich15}.  Finally, in a companion paper to the present study, \citet{Evans22} propose the presence of a reddened starburst and an AGN in the bright NE and SW cores, respectively, of VV 114E based on broadband \emph{JWST} mid-IR colors.

In this paper we present the James Webb Space Telescope (\emph{JWST}) combined MIRI/NIRSpec integral field spectroscopic observations of the nucleus of VV 114E and surrounding regions taken as part of the Early Release Science (ERS) program 1328 (Co-PIs: L. Armus and A. Evans). The data allow us to resolve the properties of the two brightest sources, the NE and SW cores, as well as star clusters and diffuse emission surrounding the eastern nucleus.

Throughout the paper we adopt a cosmology of H$_{o}$=70 km s$^{-1}$ Mpc$^{-1}$, $\Omega_{M}$=0.28, and $\Omega_{\Lambda}$=0.72. The redshift of VV 114 (z=0.0202) corresponds to an angular scale of 400 pc/1\arcsec \citep{Wright06}.

%--------------------------
\begin{figure*}
	\centering
    \includegraphics[width=0.98\textwidth]{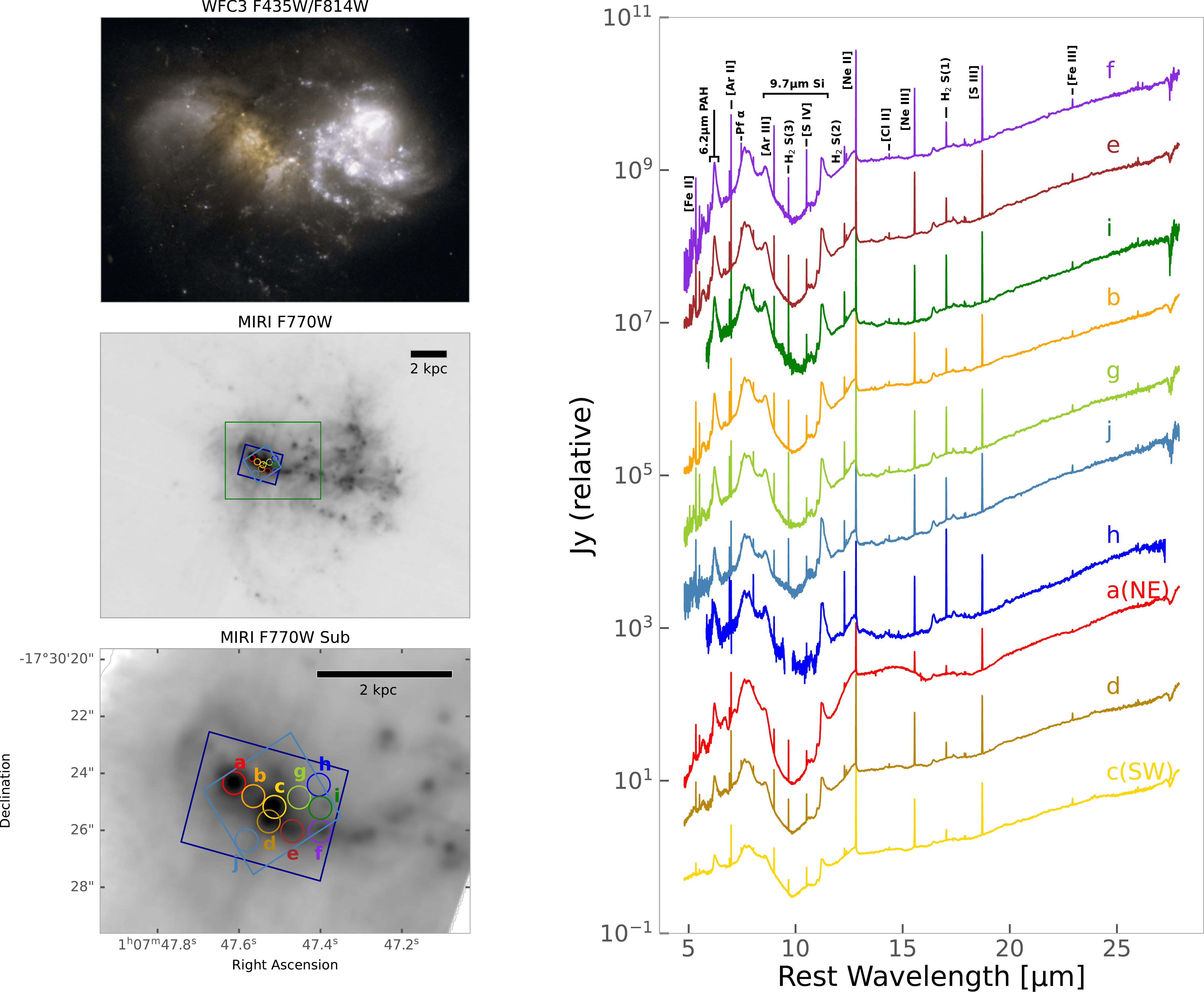}
	\caption{Images and spectra of VV 114. Top-Left: HST 435W/814W color image, Middle-Left: \emph{JWST} F770W image with the same FOV, the green box corresponds to the MIRI SUB128 subarray FOV, as shown in the Bottom-Left panel, Bottom-Left: F770W SUB128 image with  MIRI MRS channel 1 FOV (dark blue Box), NIRSpec FOV (light blue box) and our ten 0.4\arcsec~radius extraction apertures. Right: Full MIRI spectra of the 10 regions marked in the F770W subarray image. Spectra are shown at rest frame wavelengths assuming a systemic velocity of 6056 km/s and sorted from top to bottom in order of decreasing 6.2 $\mu$m PAH equivalent width.} 
	\label{FIG:FullSpectra}
\end{figure*}
%--------------------------

%--------------------------------------------
\section{Observations, Data Reduction}
%---------------------------------------------

VV 114 was observed with MIRI \cite{Rieke15,Labiano21} on 2 July 2022 (MIRI imaging, \citealt{Bouchet15}), on 5 July 2022 (MRS spectroscopy, \citealt{Wells15}), and by NIRSpec (\citealt{Jakobsen22}) on 19 July 2022 (IFU spectroscopy, \citealt{Boker22}). We include MIRI imaging in this paper to indicate the locations of our spectral extraction apertures. The full analysis of these imaging data products are described in \citet{Evans22}. 

\subsection{MIRI MRS Data}
The MIRI MRS observations include three grating settings (SHORT, MEDIUM, and LONG) in order to cover the entire wavelength range accessible to the four IFU channels (4.9-27.9$\mu$m). A four-point dither pattern was employed to recover extended emission and avoid saturation, with separate off-source observations for background subtraction. The field of view (FOV) and position angle (PA) vary by channel, but the FOV with full wavelength coverage is defined by channel 1, $\sim 5\arcsec~\times 4\arcsec$ at PA=255\textdegree~as shown in Fig. 1.

Our procedure for reducing MRS data for the ERS 1328 targets is described in detail in \citet{U22}, but a brief summary is given here: uncalibrated data are processed using the most recently available developmental release of the \emph{JWST} Science Calibration Pipeline \citep{Bushouse22}, version 1.8.3, using calibration reference data system (CRDS) context file jwst\_0963.pmap. The resulting data products generated by the pipeline are 12 background-subtracted, fringe-corrected, wavelength and flux calibrated data cubes--one for every combination of channel and grating setting combining the four dither pointings. These cubes are used to generate 1-d spectra of regions of interest and 2-d maps of the properties of particular emission line features.

We extract spectra from 10 regions of interest within the channel 1 FOV to investigate IR-bright sources resolved in the MIRI imaging as well as diffuse emission surrounding those sources and in the eastern nucleus (Fig. 1). We chose five locations coincident with the two brightest IR sources, the NE and SW cores (a, c), and several bright clumps identified in the MIRI imaging data and in both the submm and radio (d, e, f) \citep{Saito18,Evans22,Song22}. We also chose five regions intended to capture diffuse emission between (b) and around (g, h, i, j) the bright clumps coincident with some tidal and shock features previously identified using ALMA \citep{Saito17,Saito18}.

Spectra are extracted from each of the 12 data cubes using apertures with a radius of 0.4\arcsec (160 pc). This results in twelve 1-d spectra for each extracted region, with some overlap in adjacent wavelength regions. The 12 individual bands for each spectra have slight offsets in flux (a few percent) which are multiplicatively scaled, trimmed, and stitched in order to create continuously smooth 1-d spectra over the full MIRI wavelength range. This process begins by using the overlapping wavelength region to scale Channel 4 MEDIUM spectrum to the longest wavelength Channel 4 LONG spectrum ($\sim23-25\mu m$), trimming the overlapping values from the noisier spectrum of the two, and continuing the process to channels at shorter and shorter wavelengths. Finally, a wavelength dependent aperture correction is applied to each spectrum \citep{U22}.

\subsection{NIRSpec IFU Data}
NIRSpec IFU observations were taken with three grating and filter combinations: G140H/F100LP, G235H/F170LP, and G395H/F290LP to cover the spectral range from 0.97-5.3 $\mu$m. Calculations in this paper were made using the wavelength region covered by the G235H/F170LP and G395H/F290LP settings, $\sim$1.7-5.3 $\mu$m. This wavelength range allows us to measure the AGN-sensitive 3.3$\mu$m PAH feature. Again a four-point dither pattern was employed to completely sample the PSF and to avoid saturation, with an additional ``Leakcal'' image taken for each grating setting. The FOV of the combined dither pattern is $\sim3.6\arcsec~\times 3.8\arcsec$ centered on the SW Core at a PA of $\sim32^{\circ}$ (Fig. 1).

We reduce the NIRSpec data in a similar fashion to the MRS data, using calibration reference data system (CRDS) context file jwst\_1009.pmap. Uncalibrated four-point dither pattern science images and Leakcal images, one for each of the two NIRSpec chips, are downloaded using MAST resulting in 10 image files. These are first put through Stage 1 processing with the {\tt Detector1} pipeline, which applies detector-level calibrations and produces count rate files calculated from the non-destructive ``ramp'' readouts. These rate files are then processed using the {\tt Spec2} pipeline, which applies physical corrections and flux and wavelength calibrations. At this step in the overall pipeline, the Leakcal files are also used to correct for any stray light that may fall on the detector due to failed open MSA shutters. Finally we run the {\tt Spec3} pipeline step, which produces a final combined data cube sampled with $0.1\arcsec$ spaxels.

For our analysis in this paper, we extract from the final data cube spectra from the two brightest IR sources, the NE and SW cores (a and c). We matched our apertures to the MIRI MRS extraction radius of $0.4\arcsec$ centered at the same two locations. For the G395H/F290LP setting this produces flux and wavelength calibrated 1-d spectra covering 2.87-5.27$\mu$m, with a gap in coverage from 4.06-4.18$\mu$m in the middle of the spectrum due to the gap between the two NIRSpec chips. For G235H/F170LP the range is 1.66-3.17$\mu$m with a gap from 2.40-2.45$\mu$m. We use the overlapping wavelength range between the NIRSpec and MIRI data to scale the G395H/F290LP spectrum to match the MIRI spectrum, and the overlapping region between the two NIRSpec settings to scale and stitch the shorter wavelength spectra (G235H/F170LP) to the longer wavelength NIRSpec spectra.

%%-------------------------------
%LMfit results
%Full Table
%Full Table
\begin{table*}
\caption{Spectral Feature Strengths, Fluxes, and FWHM}
\centering
\begin{tabular}{cccccchhhhhhhhhhhhhhhhhhcc}
\hline
\hline
Region ID & EQW$_{3.3}$ & EQW$_{6.2}$ & s$_{9.7}$   & [Fe II] 5.34 & FWHM & [Ar II] & FWHM & [Ar III] & FWHM & [S IV] & FWHM & [Ne II] & FWHM & [Ne III] & FWHM & [SIII] & FWHM & H$_{2}$ S(5) & FWHM & H$_{2}$ S(2) & FWHM & Hu$\alpha$ & FWHM & Pf$\alpha$ & FWHM\\

\hline%\extravspace

a (NE Core) & 0.121$\pm$0.001 & 0.264$\pm$0.018 & -2.45$\pm$0.03 & 7.41$\pm$0.22 & 194$\pm$13  & 29.3$\pm$0.6  & 179$\pm$10  & 2.93$\pm$0.07 & 169$\pm$13 & 0.93$\pm$0.04 & 167$\pm$19  & 81.0$\pm$3.47 & 178$\pm$17  & 16.5$\pm$0.5  & 169$\pm$35  & 41.6$\pm$1.9  & 132$\pm$20 & 4.48$\pm$0.15 & 129$\pm$11  & 2.32$\pm$0.15 & 123$\pm$19  & 0.52$\pm$0.21 & 119$\pm$43  & 1.56$\pm$0.62 & 142$\pm$37  \\
b           &                 & 0.514$\pm$0.017 & -1.13$\pm$0.03 & 11.6$\pm$0.29 & 186$\pm$12  & 30.1$\pm$0.7  & 162$\pm$10  & 6.90$\pm$0.18 & 159$\pm$13 & 2.33$\pm$0.09 & 155$\pm$18  & 124 $\pm$6    & 166$\pm$17  & 31.0$\pm$1.1  & 166$\pm$35  & 43.9$\pm$2.6  & 115$\pm$21 & 7.05$\pm$0.13 & 136$\pm$10  & 4.98$\pm$0.12 & 131$\pm$18  & 1.07$\pm$0.11 & 151$\pm$19  & 2.83$\pm$0.58 & 148$\pm$21  \\
c (SW Core) & 0.017$\pm$0.001 & 0.106$\pm$0.002 & -1.06$\pm$0.01 & 7.49$\pm$0.40 & 178$\pm$14  & 38.0$\pm$1.2  & 147$\pm$10  & 12.5$\pm$0.6  & 156$\pm$14 & 5.01$\pm$0.29 & 165$\pm$19  & 185 $\pm$11.4 & 163$\pm$17  & 56.1$\pm$2.6  & 168$\pm$35  & 82.8$\pm$3.8  & 126$\pm$20 & 4.71$\pm$0.26 & 160$\pm$12  & 3.54$\pm$0.17 & 144$\pm$19  & 1.52$\pm$0.22 & 140$\pm$21  & 4.76$\pm$0.92 & 154$\pm$20  \\
d           &                 & 0.199$\pm$0.015 & -1.44$\pm$0.02 & 9.87$\pm$0.38 & 166$\pm$13  & 42.8$\pm$1.1  & 139$\pm$10  & 10.5$\pm$0.3  & 147$\pm$13 & 3.41$\pm$0.12 & 152$\pm$18  & 180 $\pm$8.91 & 153$\pm$17  & 43.9$\pm$1.6  & 157$\pm$35  & 64.8$\pm$3.6  & 124$\pm$20 & 6.50$\pm$0.43 & 146$\pm$12  & 3.93$\pm$0.14 & 144$\pm$18  & 1.36$\pm$0.16 & 118$\pm$18  & 4.92$\pm$0.88 & 135$\pm$18  \\
e           &                 & 0.652$\pm$0.015 & -1.17$\pm$0.04 & 7.64$\pm$0.19 & 185$\pm$12  & 19.7$\pm$0.5  & 160$\pm$10  & 4.62$\pm$0.10 & 161$\pm$13 & 1.62$\pm$0.04 & 168$\pm$18  & 84.4$\pm$3.8  & 165$\pm$17  & 22.5$\pm$0.8  & 165$\pm$35  & 36.0$\pm$1.6  & 133$\pm$20 & 3.70$\pm$0.11 & 168$\pm$11  & 3.29$\pm$0.11 & 182$\pm$18  & 0.70$\pm$0.06 & 142$\pm$17  & 2.23$\pm$0.37 & 152$\pm$18  \\
f           &                 & 0.720$\pm$0.062 & -0.90$\pm$0.04 & 3.09$\pm$0.13 & 153$\pm$13  & 13.2$\pm$0.4  & 116$\pm$10  & 7.09$\pm$0.23 & 108$\pm$13 & 2.72$\pm$0.09 & 103$\pm$18  & 60.3$\pm$3.3  & 111$\pm$17  & 14.5$\pm$0.6  & 107$\pm$35  & 27.2$\pm$2.1  & 104$\pm$21 & 2.37$\pm$0.09 & 201$\pm$11  & 2.36$\pm$0.09 & 224$\pm$19  & 0.62$\pm$0.06 & 103$\pm$17  & 2.55$\pm$0.37 & 120$\pm$13  \\  
g           &                 & 0.491$\pm$0.036 & -1.05$\pm$0.02 & 5.03$\pm$0.20 & 221$\pm$14  & 7.79$\pm$0.14 & 193$\pm$10  & 1.72$\pm$0.04 & 190$\pm$14 & 0.56$\pm$0.03 & 193$\pm$20  & 35.2$\pm$1.3  & 186$\pm$17  & 9.95$\pm$0.29 & 186$\pm$35  & 17.7$\pm$0.4  & 172$\pm$19 & 5.31$\pm$0.13 & 242$\pm$11  & 3.89$\pm$0.14 & 218$\pm$19  & 0.29$\pm$0.02 & 192$\pm$20  & 0.79$\pm$0.16 & 192$\pm$25  \\
h           &                 & 0.34$\pm$0.17   & -0.75$\pm$0.04 & 2.32$\pm$0.57 & 147$\pm$29  & 3.36$\pm$0.26 & 206$\pm$15  & 0.70$\pm$0.10 & 216$\pm$26 & 0.18$\pm$0.05 & 202$\pm$43  & 8.35$\pm$0.32 & 217$\pm$17  & 2.41$\pm$0.06 & 222$\pm$35  & 4.46$\pm$0.35 & 207$\pm$24 & 7.18$\pm$0.31 & 237$\pm$12  & 3.71$\pm$0.14 & 230$\pm$19  & 0.06$\pm$0.01 & 215$\pm$36  & 0.24$\pm$0.17 & 160$\pm$73  \\
i           &                 & 0.56$\pm$0.14   & -0.73$\pm$0.05 & 2.78$\pm$0.39 & 231$\pm$26  & 9.40$\pm$0.64 & 224$\pm$15  & 3.45$\pm$0.33 & 242$\pm$21 & 0.67$\pm$0.08 & 260$\pm$30  & 25.7$\pm$2.0  & 209$\pm$19  & 5.85$\pm$0.27 & 222$\pm$35  & 10.8$\pm$1.0  & 133$\pm$23 & 4.30$\pm$0.34 & 312$\pm$19  & 3.01$\pm$0.18 & 279$\pm$22  & 0.24$\pm$0.02 & 235$\pm$24  & 1.08$\pm$0.36 & 224$\pm$44  \\
j           &                 & 0.359$\pm$0.045 & -0.82$\pm$0.01 & 2.30$\pm$0.14 & 270$\pm$17  & 2.53$\pm$0.11 & 253$\pm$13  & 0.58$\pm$0.04 & 229$\pm$18 & 0.39$\pm$0.03 & 253$\pm$25  & 15.6$\pm$1.2  & 205$\pm$19  & 5.25$\pm$0.27 & 229$\pm$36  & 7.98$\pm$0.59 & 183$\pm$23 & 0.70$\pm$0.04 & 181$\pm$13  & 0.95$\pm$0.02 & 182$\pm$18  & 0.16$\pm$0.02 & 264$\pm$31  & 0.24$\pm$0.07 & 243$\pm$43  \\

\hline
\hline
\end{tabular}
\begin{quote}
Values of the 3.3 and 6.2$\mu$m EQW, 9.7$\mu$m silicate strength, as well as line fluxes ($10^{-18}$ W/m$^{-2}$ ) and FWHM (km s$^{-1}$, corrected for instrumental broadening) for emission line features measured in our extracted regions that were used in our analysis and discussion. This table is a subset of the total measurements, a machine readable version of the full table is available online.
\end{quote}
\end{table*}

%%-------------------------------

\begin{figure}
	\centering
	\includegraphics[width=0.49\textwidth]{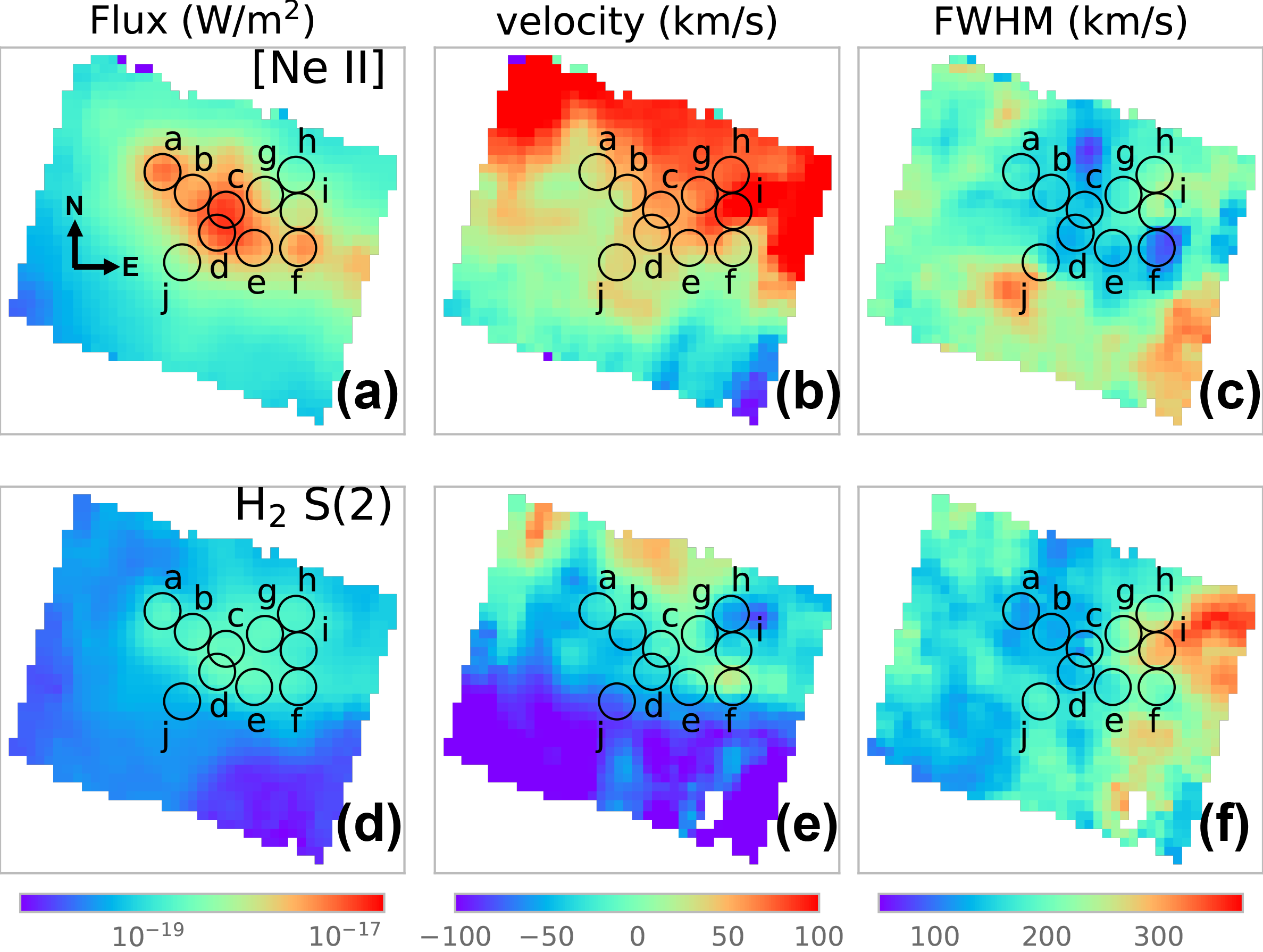}
	\caption{Two dimensional images of flux (a, d), relative velocity (b, e), and FWHM (c, f) generated from the spaxel-by-spaxel fits to the [Ne II] (12.8 $\mu$m) and H$_{2}$ S(2) (12.3 $\mu$m) emission lines. The FWHM shown is corrected for instrumental broadening as described in the text. The relative velocity map is generated by subtracting a systemic velocity of 6056 km/s from each spaxel. The broadest H$_{2}$ FWHM in the NW is just outside the Ch1 FOV but falls within the Ch3 FOV, and corresponds to the eastern edge of the shocked ``overlap'' region observed in \citet{Saito17}. The elevated FWHM in [Ne II] is near apertures that show evidence of shocks and double peaked emission line profiles (g, h, i, j). } 
	\label{FIG:Maps}
\end{figure}

%-------------------------------------------------------
\section{Results}
%-------------------------------------------------------
We use the 1-d aperture extracted spectra to measure emission and absorption features that trace the physical properties of the gas and dust. Several of our apertures are centered on bright, unresolved mid-IR sources seen in the MIRI imaging observations \citep{Evans22}, including the bright NE and SW cores (a \& c), a source directly SE of the SW core (d), and a deeply embedded star cluster (f) with Mass M$\sim 10^{6}$ M$_{\odot}$, age t $\sim 1-2$ Myr, and extinction A$_{V} \sim 8$ (see \citealt{Linden22}). The remaining apertures trace diffuse emission generally showing elevated EQW$_{6.2 \mu m}$ and strong H$_{2}$ emission.

%--------------------------
\begin{figure}
	\centering
	\includegraphics[width=0.49\textwidth]{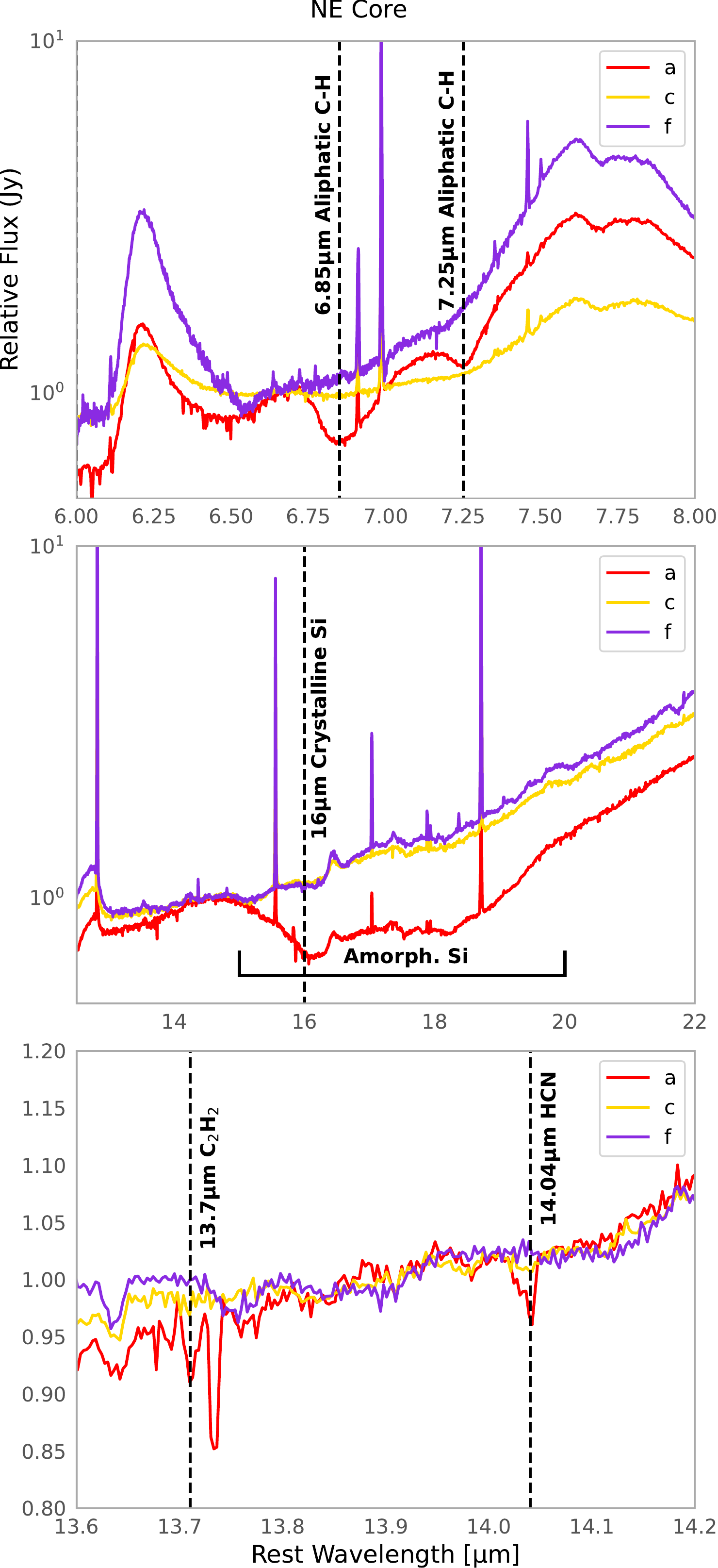}
	\caption{Expanded view of spectral features present only in the NE nucleus (region a) indicative of a highly embedded source, compared with spectra of the SW nucleus (c) and region f (embedded cluster). Dashed vertical lines in each panel correspond to the following features Top panel: Absorption features due to aliphatic hydrocarbons at 6.85 and 7.25 $\mu$m. Spectra are normalized to 1 Jy at 6.7 $\mu$m. Middle panel: crystalline silicate absorption features at 16 and 19 $\mu$m combined with amorphous silicate absorption at 18 $\mu$m. Spectra are normalized to 1 Jy at 14.5 $\mu$m. Bottom Panel: C$_{2}$H$_{2}$ and HCN absorption. Spectra are normalized to 1 Jy at 13.8 $\mu$m.} 
	\label{FIG:featuresplot}
\end{figure}
%--------------------------

\subsection{Emission Line Properties}
We perform fits to features in each 1-d MIRI spectrum using the ``{\tt lmfit}'' package \citep{Newville14} with resulting values given in Table 1. Atomic and H$_{2}$ emission lines are fit with a single Gaussian component combined with a polynomial fit to the local continuum over a range of 0.1$\mu$m. The resulting Gaussian parameters are used to derive the observed flux of each line. The width of the Gaussian fit is used to determine the intrinsic FWHM of each emission line by subtracting in quadrature the instrumental resolution of MRS at the observed wavelength \citep{Labiano21}.

Several fine structure lines and H$_{2}$ lines are well detected and resolved, as well as the hydrogen recombination lines Pfund $\alpha$ and Humphreys $\alpha$ and $\beta$. Emission line ratios show variation between apertures indicative of both widespread star formation and shock excitation. We do not detect any of the high-ionization coronal lines typically found in mid-IR spectra of AGN dominated galaxies (e.g. [Ne V]; \citealt{Genzel98,Lutz00,Sturm02,Weedman05,Armus07}) in contrast to \emph{JWST} observations of NGC 7469 that show nine well-detected coronal lines excited by high-energy photons from the central AGN \citep{Armus22}.

Emission line ratios sensitive to AGN activity (e.g [S IV]/[Ne II], [O IV]/[Ne II]) show values indicative of a composite of AGN and starburst activity \citep{Inami13}. Apertures g, h and j show elevated values of [Fe II]/Pf$\alpha$ ($>5$) and g \& j show [Fe II] FWHM higher than the other apertures (200-300 km/s).

Apertures a, b, c, d, and e have fine structure lines ([Ar II], [Ar III], [Ne II], [Cl II], [Ne III], [S IV]) with FWHM ranging between 150-200 km/s and show no trend with emission line ionization potential. In the NE core (a) the H$_{2}$ FWHM is $\sim$50 km/s narrower than the fine structure lines, while in the SW core (c) the H$_{2}$ and fine structure line widths are similar. This trend is reversed in aperture f where the H$_{2}$ FWHM are $\sim$200 km/s vs. $\sim$100 km/s for the fine structure lines. The broadest emission line widths in our data are observed in the spectrum of aperture i, which samples the diffuse H$_{2}$ bright gas near the western edge of our FOV close to the ``overlap'' region defined in \citep{Saito17}. 

In order to assess the morphology and kinematics of the region observed with MIRI/MRS, we have also carried out spaxel-by-spaxel fits of the [Ne II] 12.8 $\mu$m fine structure emission line and H$_{2}$ S(2) 12.3 $\mu$m molecular emission line. These two emission lines are generally quite luminous and are well detected across nearly every spaxel. The lines are covered by Channel 3B (13.29-15.52$\mu$m) which provides a wider field of view ($7\arcsec\times6\arcsec$) while still maintaining relatively high spectral resolution (R$\sim$2800-3000) with somewhat larger spaxels ($\sim$0.2\arcsec). Fits to the two emission lines are carried out on a spaxel-by-spaxel basis using the Channel 3B sub-band data cube, with a single Gaussian component fit to each line independently, in the same manner as the aperture extracted spectra. The resulting maps are shown in Figure 2. 

The variation in the flux and FWHM of both the [Ne II] and H$_{2}$ lines agrees with the values measured in our extracted apertures. The broadest FWHM in both lines lies outside the Channel 1 FOV, our closest aperture extractions are i and j. The increase in H$_{2}$ FWHM at the NW corner of our map corresponds to a portion of the ``overlap'' region observed in \citet{Saito17} with a similar increase in FWHM seen at submm wavelengths.

\subsection{PAH Equivalent Width}
We measure the equivalent width of the 3.3 $\mu$m and 6.2 $\mu$m PAH emission features (EQW$_{3.3\mu m}$, EQW$_{6.2\mu m}$) in the NIRSpec and MIRI spectra by applying the same method to both. First, portions of the spectrum adjacent to each PAH feature are used to perform a linear interpolation of the continuum. We then integrate a spline fit from 3.20--3.28 and 5.95--6.55 $\mu$m (rest frame, EQW$_{3.3\mu m}$, EQW$_{6.2\mu m}$) to calculate the PAH feature flux. The integrated flux is divided by the continuum flux density at the wavelength of the peak of each PAH feature.

The EQW$_{6.2\mu m}$ values range from 0.11--0.72 $\mu m$, bracketing the published \emph{Spitzer} IRS value of 0.3 $\mu m$ \citep{Stierwalt13}. The NE core has an EQW$_{6.2\mu m}$ of 0.264$\mu$m and a relatively high EQW$_{3.3\mu m}$ of 0.121$\mu$m, while the SW core has a low EQW$_{6.2\mu m}$ of 0.106$\mu$m and a very low values of EQW$_{3.3\mu m}$ of 0.016$\mu$m, indicative of an AGN (e.g. \citealt{Imanishi10,Petric11}, see discussion).

The largest EQW$_{6.2\mu m}$ of 0.72 $\mu$m is measured in source f, a very young, highly enshrouded star cluster revealed by \emph{JWST} NIRCam \citep{Linden22}. The remaining apertures have a range of values from $\sim0.20$--0.65$\mu$m tracing the presence of extended star formation and the influence of the AGN in the SW core.

\subsection{Silicate Absorption Strength}
To compare with previously published values we also calculate the apparent 9.7$\mu$m silicate absorption feature strength in a manner consistent with \citet{Spoon07} measurements of PAH-dominated sources. We assume an extrapolated power law fit for the continuum, constrained by portions of the spectrum at 5.5$\mu$m and 14$\mu$m and take the natural logarithm of the ratio of the observed and interpolated continuum flux density at 9.7$\mu$m (s$_{9.7\mu m}$). 

The majority of the s$_{9.7\mu m}$ values measured range from -0.73 to -1.44, bracketing the published \emph{Spitzer} IRS measurement of -0.98, with the exception of the NE core (a). Visual inspection of the NE core's spectrum indicates a much deeper silicate absorption feature, measured to be s$_{9.7\mu m}$=-2.45. For absorption dominated sources \citet{Spoon07} used a spline fit to determine a continuum. If we assume the NE core is absorption dominated and apply the same process, the measured s$_{9.7\mu m}$ would be stronger (-2.87). Several other features unique to the spectrum of the NE source are shown in Fig. 3 including aliphatic hydrocarbon absorption and crystalline silicate absorption, both observed in ULIRGs with deep 9.7$\mu$m silicate absorption \citep{Spoon07}. We note that these strong absorption features are not seen in either the SW (c) core or the embedded cluster (f).

\subsection{Molecular Absorption Features}
The MIRI spectra allow high resolution vibration$-$rotation spectroscopy of gas-phase molecules towards dust-enshrouded regions in the VV 114 system. We report the detection in absorption of the $\nu_2$ 14.04 $\mu$m bending mode of hydrogen cyanide (HCN) and tentatively also the $\nu_5$ 13.7 $\mu$m bending mode of acetylene (C$_2$H$_2$) towards the NE Core (a). C$_2$H$_2$ is a key ingredient in the gas-phase formation of large molecules such as HC$_3$N, and HCN is one of the most abundant nitrogen bearing molecules in dense ($n>1 \times 10^4$ cm$^{-3}$) molecular clouds.

These lines have been previously detected by {\it Spitzer} towards dust enshrouded young stellar objects (YSOs) (e.g. \citealt{Lahuis00}) and towards luminous and obscured infrared galaxies (LIRGs) \citep{Lahuis07}. \citet{Lahuis07} find HCN column densities ranging between  $N$(HCN)=$1-12 \times 10^{16}$ cm$^{-2}$ and warm gas with excitation temperatures $T_{\rm ex}$=230-700 K.  Owing to the higher spatial and spectral resolution,  the JWST MIRI spectrum of the NE core is more complex than those found by \citet{Lahuis07} using {\it Spitzer}.

We perform preliminary fits using the methodology of \citet{Lahuis00} which assumes LTE. Our preliminary results are consistent with an excitation temperature of 300-500 K and an $N$(HCN) of $1-5 \times10^{16}$ cm$^{-2}$. For an HCN abundance (with respect to H$_2$) of $10^{-8}$-$10^{-7}$ (e.g. \citealt{Schilke92,Lahuis00,Harada13}), this would be consistent with a high obscuration with $N$(H$_2$) $10^{23}$-$10^{24}$ cm$^{-2}$. The HCN and C$_2$H$_2$ spectra require further analysis, including multiple temperature component modeling and inclusion of non-LTE effects (Buiten et al. in prep).

It is also possible that the nuclear obscuration is even higher with column densities in excess of $N$(H$_2$) $10^{25}$ cm$^{-2}$--the so called Compact Obscured Nuclei (CONs, e.g. \citealt{Aalto15}). Such objects are characterized by high mm and submm continuum surface brightness and luminous emission from mm-wave rotational transitions of HCN within the vibrationally excited  ladder (HCN-vib) (e.g. \citealt{Sakamoto10,Aalto15,Aalto19,Sakamoto21,Falstad21}). 
For such deeply enshrouded objects, the HCN 14 $\mu$m line, or the silicate absorption, may not trace the full $N$(H$_2$), but only its surface. The mm/submm HCN-vib line (and the mm/submm continuum) would probe deeper, revealing the full obscuring column.
\citet{Falstad21} propose that CONs have surface brightness of HCN-vib of $\Sigma$(HCN-vib)$>$ 1 L$_{\odot} pc^{-2}$. A recent ALMA study by \citet{Saito18} does not detect HCN-vib emission towards the NE Core of VV 114 with a limit of $\Sigma$(HCN-vib)$<$ 0.12 L$_{\odot} pc^{-2}$. The relatively faint mm continuum found by \citet{Saito17} is consistent with the HCN-vib non-detection. This suggests that the NE Core either does not fulfil the CON criteria of \citet{Falstad21} or that the radius of the CON region is smaller than 12 pc.

%--------------------------
\begin{figure*}
	\centering
	\includegraphics[width=0.98\textwidth]{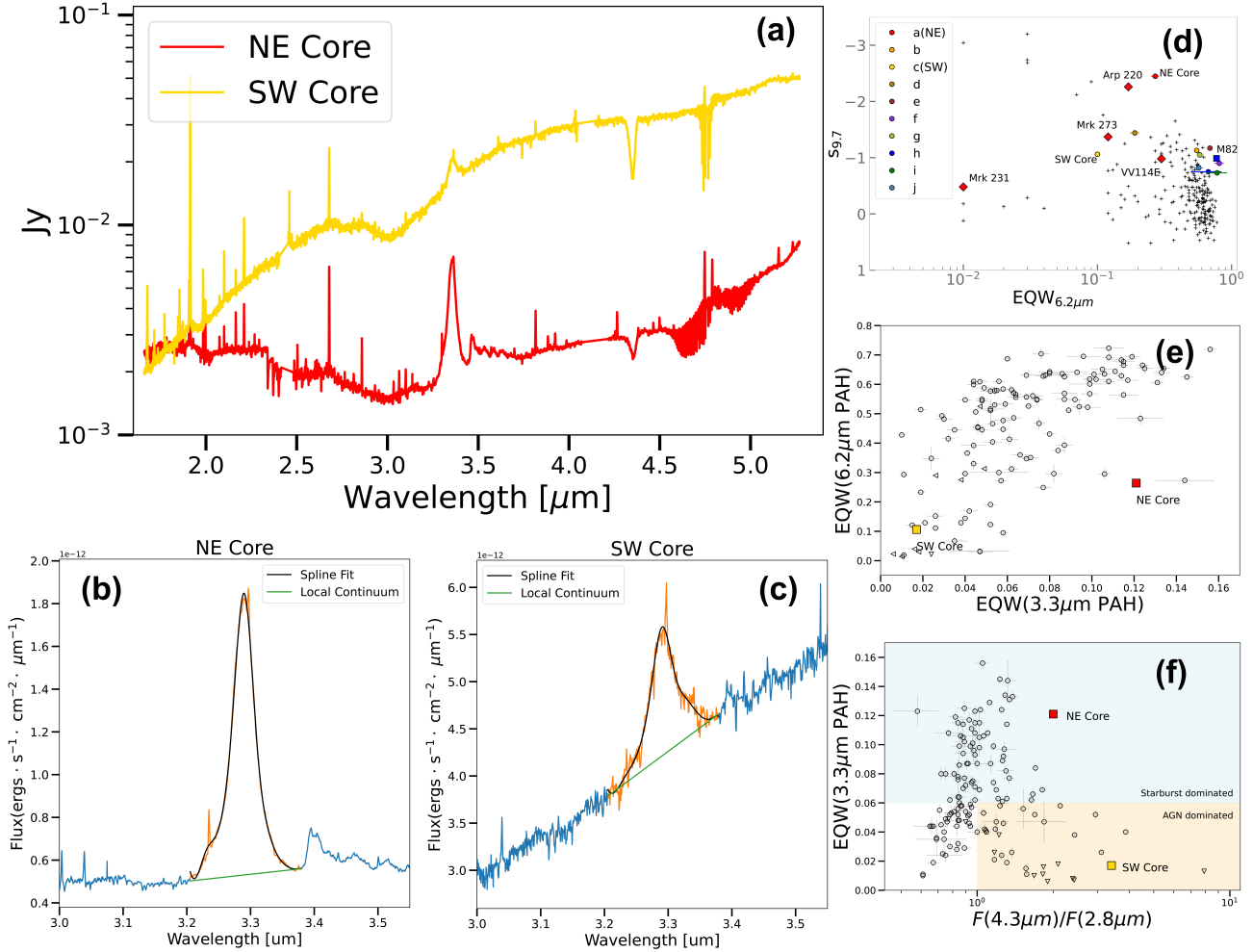}
	\caption{NIRSpec spectra and PAH diagnostic plots used (a): NIRSpec Spectra of the NE and SW cores in VV 114. The SW core shows a strongly rising continuum characteristic of hot dust associated with an AGN. The 3.3 $\mu$m PAH feature is detected in both spectra, but is significantly weaker in the SW core. (b) \& (c): expanded region and fit to the 3.3 $\mu$m PAH feature in the NE (b) and SW (c) cores. (d): Equivalent width of the 6.2$\mu$m PAH feature plotted vs. Silicate strength. Measurements for our MIRI apertures plotted as colored circles. Values measured with \emph{Spitzer}/IRS for ULIRGs and LIRGs from the GOALS sample are plotted as ``+'' symbols, with VV 114E and three comparative ULIRGs denoted with red diamonds \citep{Stierwalt13}. Starburst galaxy M82 is shown as a blue square \citep{Spoon07}. (e): Equivalent widths of 6.2$\mu$m and 3.3$\mu$m PAH features for the SW and NE cores \citep{Inami18}. (f): Flux Density ratio and 3.3$\mu$m PAH EQW for the SW and NE cores \citep{Inami18}. The SW core lies near AGN-dominated ULIRGs in both diagnostic plots.} 
	\label{FIG:Diagnostics}
\end{figure*}
%--------------------------

%------------------
\section{Discussion}
%------------------
The \emph{JWST} spectra and imaging resolve a blend of diffuse emission from star formation and shocks, several reddened star-forming knots, and an AGN. Although the resolved spectra cover a much smaller region, our results confirm the analysis of MIRI imaging data in \citet{Evans22}. The SW core (c) has an elevated continuum at $\sim5\mu$m, consistent with the presence of dust in thermal equilibrium at temperatures near the sublimation temperature of the silicate grains ($\sim$1200 K). This elevated continuum has been demonstrated by \citet{Laurent00,Petric11} (and references therein) as a telltale sign of a dust enshrouded, optically thick AGN. We also find PAH emission in all of our extracted apertures and widespread diffuse emission in our emission line maps. The various emission and absorption features allow us to directly probe the nature of the bright cores and diffuse emission as well as the kinematics of the gas in VV 114E.
 
\subsection{Evidence for AGN Activity}
Previous multiwavelength observations of VV 114 hinted at an AGN contribution in the X-ray, mid-IR, and submm (e.g. \citealt{Lefloch02,Grimes06,Iono13}). As the brightest compact sources from the mid-IR to radio, the NE and SW cores are the most likely to harbor an AGN, with \citet{Iono13} identifying the NE core as a potential AGN. We do not detect coronal lines in either the NE or SW core, but this does not rule out an AGN; the ULIRG Mrk 231 is a well-known optically classified Seyfert 1 with no observed coronal lines in the mid-IR. Instead, very low PAH equivalent width and silicate absorption indicate the presence of the AGN in Mrk 231 \citep{Armus07, Imanishi10, Inami13, Stierwalt14}.

To compare with other AGN and starburst dominated LIRGs we plot our measured values on several PAH diagnostic diagrams (Fig. 4). We first plot s$_{9.7}$ vs. EQW$_{6.2\mu m}$ (Fig. 4d): the apertures extracted from diffuse emission (with the exception of d) as well as the star forming clump in aperture f have values consistent with starburst galaxies and lie near the literature values for M82 \citep{Spoon07}. The SW core shows contribution from an AGN (e.g. \citealt{Spoon07,Marshall18}) and lies near Mrk 273, a ULIRG with a Seyfert 2 nucleus \citep{Armus07,Stierwalt13,U13}. Aperture d is directly adjacent to the SW core, in fact partially overlapping, and is likely showing some contribution from the AGN that is more clearly seen in the SW core. The NE core has a correspondingly higher obscuration and lies near the values observed for the ULIRG Arp 220, a deeply embedded starburst that also has C$_{2}$H$_{2}$ and HCN absorption crystalline silicate features in its mid-IR spectrum \citep{Spoon06,Lahuis07}.

The spectrum of the SW core at shorter IR wavelengths is consistent with the heating and processing of dust by an AGN, which reduces the 3.3$\mu$m PAH EQW (Fig. 4a-c) as well as the 6.2$\mu$m EQW (EQW$_{3.3}<0.04\mu$m and EQW$_{6.2}<0.20\mu$m, \citealt{Imanishi10,Petric11}). The 3.3$\mu$m PAH feature is directly adjacent to a broad absorption feature caused by H$_{2}$O ice that may impact our continuum measurement and in turn the measured EQW$_{3.3\mu m}$ \citep{Imanishi08,Inami18}. Moreover, the attenuation of the continuum and the 3.3$\mu$m feature may be different depending on the geometry of the dust and PAH emisison \citep{Lai20}. We make a simple estimate of the impact by performing a power law fit to the 1.7-5.0$\mu$m spectrum and divide our integrated flux by the continuum flux density at 3.3$\mu$m. This decreases the EQW$_{3.3\mu m}$ slightly from 0.12$\mu$m and 0.017$\mu$m to 0.11$\mu$m and 0.015$\mu$m for the NE and SW core, but does not affect our conclusions regarding the nature the NE and SW cores.

\citet{Inami18} used AKARI to propose revised AGN diagnostics including the 3.3 $\mu$m PAH feature as well as the $F_{\nu}(4.3)/F_{\nu}(2.8)$ flux density ratio. When placed on a plot of EQW$_{6.2\mu m}$ vs. EQW$_{3.3\mu m}$ (Fig. 4e, following \citealt{Inami18}), the SW core lies in a region populated by known AGN while the high EQW$_{3.3\mu m}$ of the NE core is more consistent with ULIRGs dominated by star formation. We also place the NE and SW core on the EQW$_{3.3}\mu$m vs. flux density ratio diagnostic proposed by \citet{Inami18} in Fig. 4f. The SW core again clearly lies in the AGN-dominated portion of the diagram. 

To estimate the contribution of the AGN in the SW core to the total luminosity of VV 114 we take the estimation of L$_{IR}\sim5\pm0.5\times10^{10}L_{\odot}$ by \citet{Evans22}, about 12\% of the total L$_{IR}$, as an upper limit. The presence of PAH features in the SW core indicates a blended contribution to the IR of star formation and AGN. Sources with similar EQW$_{3.3}$ and colors have a bolometric AGN contribution of $\sim30-50\%$ \citep{DiazSantos17, Inami18}, which means the contribution of the AGN to the total luminosity of the entire VV 114 system is $\sim5\%$. 

Although \citet{Iono13} identified the NE core as potentially harboring an AGN due to the enhanced HCN/HCO+ ratio, an analysis of X-ray and millimeter data by \citet{Privon20} showed that in fact the HCN/HCO+ is not a robust indicator of total AGN luminosity or its fractional contribution to infrared luminosity. Our findings regarding the nature of the two bright cores in VV 114E are not in agreement with the analysis of \emph{JWST} MIRI data by \citet{Donnan22} who propose the NE core as a potential AGN host based on its compact nature.

Interestingly, the NE core does have a somewhat low EQW$_{6.2\mu m}$ despite a strong EQW$_{3.3\mu m}$, placing it in a region of Fig. 4e with few other galaxies. The galaxies with the most similar values to the NE core are II Zw 96 and IRAS F19297-0406, which also have similar s$_{9.7 \mu m}$ when comparing the values measured using \emph{Spitzer} for all three galaxies \citep{Stierwalt13}. II Zw 96 is an unusually compact and powerful starburst (\citealt{Inami10,Inami22}) and IRAS F19297-0406 has a powerful starburst driven outflow (\citealt{Soto12,Veilleux13}), physical conditions which may warm dust in a way that lowers EQW$_{6.2\mu m}$. The curious nature of both the NE and SW cores warrants a more thorough follow-up analysis decomposing the relative contribution of the stellar, PAH, dust, and AGN components of the SED.

\subsection{Shocks and Tidal Features}
Evidence for extended shocks in VV 114 has previously been suggested at visible wavelengths via enhanced [O I] and [S II] emission and broadened line profiles across VV 114 \citep{Rich11,Rich15} and in the submm from the presence of methanol (CH$_{3}$OH) in the ``overlap'' region between VV 114W \& E. Apertures f, g, h, and i lie closest to the overlap region and both emission line ratios and molecular and atomic line widths show evidence of shocked gas. 

The atomic lines in aperture f all have narrow line widths ($\sim$100-150 km/s) and ratios typical of buried, young star-forming regions as this source was revealed to be in \citep{Linden22}. Aperture i encompasses a fainter star forming knot seen in both MIRI and NIRCAM imaging, and several emission line profiles show a double peaked profile. The broader H$_{2}$ lines in apertures f and i ($\sim$200-300 km/s) are similar to the values extracted just to the north in apertures g and h, which show elevated [Fe II]/Pf$\alpha$ values indicative of shocks, also seen in MRS observations of NGC 7469 \citep{U22}. 

These regions are consistent with ALMA observations of VV 114 where the elevated FWHM of 150-300 km/s in the molecular gas (e.g. CO(1-0), CH$_{3}$OH) is found in the ``overlap'' region and is suggested to be the product of both shocks, and overlapping kinematic components due to the merger \citep{Saito17,Saito18} and may be similar in nature to the shocked bridge seen in Stephan's Quintet  \citep{Guillard12b,Appleton17}. However, the clusters in this region of VV 114 are 1-2 orders of magnitude more massive and $\sim$2-3 times dustier than those seen in the bridge of Stephan's Quintet \citep{Fedotov11,Linden22}.

Resolved emission line profiles in aperture j show a strong indication of both blue and red-shifted wings, potentially due to projection effects of the tidal arm that extends from VV 114W across the IR bright cores southward beyond the FOV of our aperture extractions \citep{Saito15}. These values are supported by examining the [Ne II] emission line profiles which fall in Channel 3 and has a wider FOV than Channel 1. Looking several arcseconds southeast of the SW core, the [Ne II] line profiles show significant broadening and wings in several spaxels (Fig. 2). We also see enhanced H$_{2}$/[Ne II] in this region and at the furthest SE region of our spaxel-by-spaxel maps. The value of [Fe II]/Pf$\alpha\sim9$ in aperture j also indicates the presence of shocked gas which likely extends beyond the Channel 1 FOV and follows the elevated emission line ratios caused by shocks seen in visible light IFU observations \citep{Rich11}.

Aperture b appears to be dominated by diffuse emission when examining the MIRI and NIRCAM images, but does not display the same characteristics of shock excitation as the other apertures that trace diffuse gas. Previous observations of shock excitation in VV 114 have suggested both galactic winds and tidally driven gas flows as sources of shock excitation \citep{Rich11, Saito17}. The \emph{JWST} data show some shocked gas coincident with tidal features previously observed in the submm and radio as well as kinematic profiles potentially associated with galactic winds. Follow-up work mapping the two dimensional kinematics of the molecular and atomic gas, as well as the temperature and distribution of H$_{2}$ gas using these data will provide a more complete picture of the shock excitation in VV 114, especially when combined with wide field optical IFU observations from MUSE.

\subsection{Star Formation Rates}
If we assume that the atomic and fine structure line emission is dominated by star formation in the apertures with bright star-forming clumps (a, e, f), we can calculate star formation rates using a hydrogen recombination line flux or Neon emission. If we scale either Pfund or Humphreys $\alpha$ to H$\alpha$ assuming Case B recombination (Hummer \& Storey 1987) and use equation (2) in \citet{Murphy11} the estimated SFR for each spectrum is $\sim1$M$_{\odot}$/yr. Using the [Ne II] and [Ne III] fluxes and equation (3) in \citet{Ho07} yields an SFR from $\sim1.5-2.5$M$_{\odot}$/yr. These values are consistent with those found by \citet{Song22} using the radio emission of bright knots measured with the VLA and amount to $\sim2-3\%$ of the total SFR per aperture.

%--------------------------------------------
\section{Conclusions}
%--------------------------------------------
Our analysis of MIRI MRS and NIRSpec IFU spectroscopy of VV 114E shows emission and absorption features that allow us to resolve variations in the properties of the IR bright nuclear cores, unresolved clumps, and diffuse gas. The integrated properties of the two bright cores and the diffuse gas agree with past multi-wavelength observations of VV 114E that show widespread star formation and diffuse emission from shocks, and reveal spectroscopic evidence of an AGN in the SW core. More specifically:

\begin{itemize}
\item
The SW core harbors an AGN as indicated by its extremely low 3.3 and 6.2 $\mu$m PAH equivalent widths and strong 3-5$\mu$m continuum, consistent with AGN-dominated LIRGs. The SW core is also surrounded by star forming knots and diffuse emission, which is evident in the atomic line ratios at longer wavelengths. The AGN in the SW core likely accounts for $\sim5\%$ of the total luminosity of VV 114.

\item 
The NE core is deeply embedded, its mid-IR spectrum displays strong 9.7$\mu$m silicate absorption, crystalline silicate features, aliphatic hydrocarbons, and HCN absorption. Using the 14$\mu$m HCN absorption line we calculate a temperature and column density of 300-500 K and $N$(HCN) of $1-5 \times10^{16}$ cm$^{-2}$. Our data show no evidence of an AGN impacting the atomic or molecular gas at mid-IR wavelengths and we conclude that this source is a deeply buried star forming region.

\item 
The diffuse gas NW of the nuclear region shows elevated [Fe II]/Pf$\alpha$ and higher H$_{2}$ line widths along with double peaked profiles caused by shocked gas in the overlap region previously observed by $ALMA$. A fit to [Ne II] across the MIRI FOV reveals broader emission profiles to the south of the nucleus consistent with the extended tidal feature observed by $ALMA$, as well as shocked gas $\sim$1.5\arcsec~SE of the SW Core that likely extends beyond the Channel 1 FOV, consistent with previous visible light IFU observations.

\end{itemize}

These early 3D spectral data highlight the power of combining the NIRSpec and MIRI data to elucidate the nature of complex, obscured star formation and AGN in the local Universe. Taken together the spectroscopic datasets from both \emph{JWST} instruments are extremely rich and will facilitate detailed and thorough analysis in future papers in this series.

%--------------------------------------------
\section*{Acknowledgements}
%--------------------------------------------
We thank the referee for their helpful comments. This work is based on observations made with the NASA/ESA/CSA \emph{JWST}. The research was supported by NASA grant JWST-ERS-01328. The data were obtained from the Mikulski Archive for Space Telescopes at the Space Telescope Science Institute, which is operated by the Association of Universities for Research in Astronomy, Inc., under NASA contract NAS 5-03127 for JWST. The specific observations analyzed can be accessed via \dataset[10.17909/yqk1-jr92]{https://doi.org/110.17909/yqk1-jr92}. VU acknowledges funding support from NASA Astrophysics Data Analysis Program (ADAP) grant 80NSSC20K0450. The Flatiron Institute is supported by the Simons Foundation. AMM acknowledges support from the National Science Foundation under Grant No. 2009416. ASE and SL acknowledge support from NASA grants HST-GO15472 and HST-GO16914. YS was funded in part by the NSF through the Grote Reber Fellowship Program administered by Associated Universities, Inc./National Radio Astronomy Observatory. The National Radio Astronomy Observatory is a facility of the National Science Foundation operated under cooperative agreement by Associated Universities, Inc. F.M-S. acknowledges support from NASA through ADAP award 80NSSC19K1096. SA gratefully acknowledges support from an ERC Advanced Grant 789410, from the Swedish Research Council and from the Knut and Alice Wallenberg (KAW) Foundation. SA gratefully acknowledges John Black for helpful discussions. KI acknowledges support by the Spanish MCIN under grant PID2019-105510GB-C33/AEI/10.13039/501100011033. HI and TB acknowledge support from JSPS KAKENHI grant No. JP21H01129 and the Ito Foundation for Promotion of Science. This work was also partly supported by the Spanish program Unidad de Excelencia Mar'a de Maeztu CEX2020-001058-M, financed by MCIN/AEI/10.13039/501100011033. The computations presented here were conducted through Carnegie's partnership in the Resnick High Performance Computing Center, a facility supported by Resnick Sustainability Institute at the California Institute of Technology. Finally, this research has made use of the NASA/IPAC Extragalactic Database (NED) which is operated by the Jet Propulsion Laboratory, California Institute of Technology, under contract with the National Aeronautics and Space Administration.

\vspace{5mm}

\facilities{\emph{JWST} (NIRCam, NIRSpec and MIRI)}%, Keck, ALMA, HST}

%Hello people on Twitter, hope you're having a lovely day.

%--------------------------------------------
%--------------------------------------------

\bibliographystyle{apj}
\bibliography{ms}

%--------------------------------------------

%--------------------------------------------

\end{document}